# Plagiarism Detection Using Graph-Based Representation

[1]Ahmed Hamza Osman, [2]Naomie Salim, [3]Mohammed Salem Binwahlan

[1]Faculty of Computer Science, International University of Africa, Sudan
[2]Faculty of Computer Science and Information Systems, Universiti Teknologi Malaysia
[3]Faculty of Applied Sciences, Hadhramout University of Science & Technology, Yemen

[1, 2, 3] 81310, Skudai, Johor, Malaysia
Tel +60147747409/+607 5532208   Fax   +607 5532210

**Abstract**—Plagiarism of material from the Internet is a widespread and growing problem. Several methods used to detect the plagiarism and similarity between the source document and suspected documents such as fingerprint based on character or n-gram. In this paper, we discussed a new method to detect the plagiarism based on graph representation; however, Preprocessing for each document is required such as breaking down the document into its constituent sentences. Segmentation of each sentence into separated terms and stop word removal. We build the graph by grouping each sentence terms in one node, the resulted nodes are connected to each other based on order of sentence within the document, all nodes in graph are also connected to top level node" Topic Signature ". Topic signature node is formed by extracting the concepts of each sentence terms and grouping them in such node. The main advantage of the proposed method is the topic signature which is main entry for the graph is used as quick guide to the relevant nodes. which should be considered for the comparison between source documents and suspected one.  We believe the proposed method can achieve a good performance in terms of effectiveness and efficiency.

——————————— ◆ ———————————

## 1 INTRODUCTION

PLAGIRAISM is one of the forms of misuse of academic activities has increased rapidly in the quick and easy access to data and information through electronic documents and the Internet, and when we talk about plagiarism, we mean the text written by others where they are re-adjust the text to format by adding or deleting without any citation or reference.

There are many types of plagiarism, such as copy and paste, which is the most common, redrafting or paraphrasing of the text, plagiarism of the idea, plagiarism through translation from one language to another and many other methods that use plagiarism. Plagiarism is a serious problem in computer science. This is partly due to the ease with which electronic assignments may be copied, and to the difficulty in detecting similar assignments in a sufficiently large class. In addition, students are becoming more comfortable with cheating. A recent study found that 70% of students admit to some plagiarism, with about half being guilty of a serious cheating offense on a written assignment. In addition, 40% of students admit to using the "cut-and-paste" approach when completing their assignments [1]. The key and main issue in plagiarism detection field is how to differentiate between plagiarized document and non-plagiarized document in effective and efficient way.

The current methods of plagiarism detection relay on the comparison of small text unit such as character, n-gram, chunk or terms. Suppose we have a document contents ten sentence, each sentence contains five terms and each term consists of at least one character. The consideration of small text unit (character) for detecting of similarity between the original document and suspected document lead to a huge number of comparisons. In this paper, we propose a new method for plagiarism detection. The proposed method is graph-based, where each document is represented as graph. One node represents one sentence. Top level node is different node where it contains the concepts of terms in the document. Such node is called topic signature. The main advantage of proposed method is the topic signature which is the main entry for the graph is used as quick guide to the relevant nodes, which should be considered for the comparison between source documents and suspected one.

————————————————

- *Ahmed Hamza Osman. is with the Faculty of Computer Science, International University of Africa, Sudan.*
- *Naomie Salim. Faculty of Computer Science and Information Systems, Universiti Teknologi Malaysia,malsysia.*
- *Mohammed Salem Binwahlan. is with the Faculty of Applied Sciences, Hadhramout University of Science & Technology, Yemen.*

This paper is structured as follows. Section 2 introduces the current state of the studies on plagiarism detection. While Section 3 presents an overview of the graph based representation, Section 4 describes the concepts extraction for the sentence. Contribution and comparison methods are reported in Section 5. Finally, Section 6 concludes this paper.

## 2 RELATED WORK

### 2.1 Plagiarism Detection Overview

In plagiarism detection, a correct selection of text features in order to discriminate plagiarised from non-plagiarised documents is a key aspect. [2] Has delimited a set of features which can be used in order to find plagiarism cases such as changes in the vocabulary, amount of similarity among texts or frequency of words. This type of features has produced diferent approaches to this task. Substantive plagiarism analysis [3] is a different task from plagiarism detection with reference. It captures the style across a suspected document in order to find fragments that are plagiarism candidates. This approach saves the cost of the comparison process, but it does not give any hint about the possible source of the potentially plagiarised text fragments. In those cases where a reference corpus is considered, the search process has been based on different features. [4] Considers text comparison based on word n-grams. The reference, as well as the suspected text, is split into trigrams, composing two sets which are compared. The amount of common trigrams is considered in order to detect potential plagiarism cases. [5] Considers the sentence as the comparison unit in order to compare local similarity. It differentiates among exact copy of sentences, word insertion, word removal and rewording on the basis of a Wordnet-based word expansion process.

some authors [6] define plagiarism as "unacknowledged copying of documents or programs" that can "occur in many contexts: in industry a company may seek competitive advantage; in academia academics may seek to publish their research in advance of their colleagues." Most empirical study and analysis has been undertaken by the academic community to deal with student plagiarism, although methods of detection have found their way into the commercial world, e.g. Measuring software reuse and identifying reused code (see, e.g. [7]).

There are several schemes to characterize documents before applying one of the plagiarism detection techniques. Some document descriptors such as Character-based representation, the simplest form, in which documents are represented as a sequence of characters with ignoring spaces between words, periods (full stops) between statements and lines. Also Word-based representation, in which documents are represented as a collection of words with ignoring periods (full stops) between statements and lines. Moreover Phrase-based representation, in which a phrase (part of a statement) is used as a unit of comparison. For example, 3-word phrase or so-called trigrams can be used as a comparison unit. and Sentence-based representation, in which documents are segmented into statements using periods (full stop) as a statement-end indicator. Although Paragraphed-based representation, in which documents are described as a collection of paragraphs or passages.

There are several techniques have been developed or adapted for plagiarism detection in natural language documents. They can be classified into four main approaches. The first technique is Fingerprint Matching [8][9][10] which involves the process of scanning and examining the fingerprints of two documents in order to detect plagiarism. Then, Clustering [11][10] that uses specific words (or keywords) to find similar clusters between documents Fingerprinting techniques mostly rely on the use of K-grams [12] because the process of fingerprinting divides the document into grams of certain length k. Then, the fingerprints of two documents can be compared in order to detect plagiarism. It can, therefore, be classified fingerprints into three categories: character-based fingerprints, phrase-based fingerprints and statement-based fingerprints. The early fingerprinting technique uses sequence of characters to form the fingerprint for the whole document.

### 2.2 Plagiarism Detection Tools

Some authors refer about the tools in plagiarism detection [13] which are currently particularly popular and describe their main features in what follows.

www.plagiarism.org

Turnitin: This is a product from iParadigms. It is a web based service. Detection and processing is done remotely. The user uploads the suspected document to the system database. The system creates a complete fingerprint of the document and stores it. Proprietary algorithms are used to query the three main sources: one is the current and extensively indexed archive of Internet with approximately 4.5 billion pages, books and journals in the ProQuest™ database; and 10 million documents already submitted to the Turnitin database.

www.urkund.com

Urkund: Another server based plagiarism detection web service which offers an integrated and automated solution for plagiarism detection. It utilizes standard email systems for submission of documents and viewing results. This tool also claims to search through all available online sources giving priority to educational and scandinavian origin. This system claims to process 300 different types of document submissions.

www.copycatchgold.com

Copycatch: A client based tool used to compare locally available databases of documents. It offers 'gold' and 'campus versions', giving comparison capabilities for large number of local r sources. It also offers a web version which extends the capabilities of plagiar-ism detection across the internet using the Goggle API.

www.plagiarism.phys.virginia.edu

WCopyfind: An open source tool for detecting words or phrases of defined length within a local repository of documents . The product is being modified to extend searching capabilities across the internet net using the Google API at ACT labs10.

www.canexus.com

Eve2 (Essay Verification Engine): This tool works at the client side and uses it own internet search mechanism to find out about plagiarized contents in a suspected document.

http://www.plagiarism.com

GPSP - Glatt Plagiarism Screening Program: This software works locally and uses an approach to plagiarism detection that differs from previously mentioned services. GPSP detection is based on writing styles and pa terns. The author of a suspected submission has to go through a test of filling blank spaces in the writing. The number of correctly filled spaces and the time taken for completion of the test provides the hypothesis of plagiarism guilt or innocence.

www.cs.berkeley.edu

MOSS - a Measure of Software Similarity: MOSS Internet service "accepts batches of documents and returns a set of HTML pages showing where significant sections of a pair of documents are very similar " [14]. The service specializes in detecting plagiarism in C, C++, Java, Pascal, Ada, ML, Lisp, or Scheme programs.

www.ipd.uni-karlsruhe.de

JPlag: Another internet based service which is used to detect similarities among program source codes. Users upload the files to be compared and the system presents a report identifying matches. JPlag does programming language syntax and structure aware analysis to find results.

## 3 GRAPH DOCUMENT REPRESENTATION

Graph-based method is introduced, designed especially for web document representation [15]. " The main advantage of graph-based techniques is that they allow keeping the inherent structural information of the original document. Before describing the graph-based methodology, the definition of a graph, subgraph and graph isomorphism should be given. A graph G is a 4-tuple: $G=(V,E,\alpha,\beta)$, where V is a set of nodes (vertices), $E \subseteq V \times V$ is a set of edges connecting the nodes, $\alpha : V \rightarrow \sum v$ is a function labeling the nodes, and $\beta : V \times V \rightarrow \sum e$ is a function labeling the edges ($\sum v$ and $\sum e$ being the sets of labels that can appear on the nodes and edges, respectively). For brevity, we may refer to G as $G= (V, E)$ by omitting the labeling functions. A graph $G1=(V1,E1,\alpha1,\beta1)$ is a subgraph of a graph $G2=(V2,E2,\alpha2,\beta2)$, denoted $G1 \subseteq G2$, if $V1 \subseteq V2$, $E1 \subseteq E2$ $(V1 \times V1)$, $\alpha1(x) = \alpha2(x) \forall x \in V1$ and $\beta1(x, y) = \beta2(x, y) \forall (x, y) \in E1$. Conversely, graph G2 is also called a supergraph of G1. All graph representations proposed in [ 15] are based on the adjacency of terms in an HTML document. Under the standard method[16] each unique term (word) appearing in the document, except for stop words such as "the", "of", and "and" which convey little information, becomes a vertex in the graph representing that document. Each node is labeled with the term it represents. Note that we create only a single vertex for each word even if a word appears more than once in the text to build the terms graph in the sentence, and we create also a single vertex for each sentence this vertex involved graph of terms. Also Under the n-distance representation, there is a user-provided parameter.Instead of considering only terms immediately following a given term in a web document. For example[16], if we had the following text on a web page, "AAABBBCCCDDD", then we would have an edge from term AAA to term BBB labeled with a 1, an edge from term AAA to term CCC labeled 2, and so on. Similar to n-distance, we also have the fourth graph representation, n-simple distance. This is identical to n-distance, but the edges are not labeled, which means we only know that the distance between two connected terms is not more than n. frequency representation model is a type of graph representation too. Each node and edge are labeled with an additional frequency measure. For nodes this indicates how many times the associated term appeared in the web document; for edges, this indicates the number of times the two connected terms appeared adjacent to each other in the specified order". We discussed the representation of graphs in this paper in section 5.

## 4 CONCEPT EXTRACTION

Concept identification is a common to applications such as ontology learning, glossary extraction and keyword extraction. These applications have different definitions for concept, hence different methods. Previous methods start from the idea that concepts can be found as word or phrases contained in sentences, which are then divided into smaller phrases in one of two ways: Using grammatical or syntactical information. The former can be found in ontology learning [17], glossary extraction [18] and information retrieval systems [19]. Using a shallow grammar parser, an entire sentence is parsed into a grammatical tree, which classifies sub-phrases as noun or verb phrases, noun phrases are selected as concepts. The syntactical information division of sentences uses punctuation or conjunctions to separate phrases within a sentence, all these phrases are concepts. This approach can be found in keyword extraction systems [20].

The idea that was used in this paper refers to extraction the general concepts from the sentences by any me-thod used to extraction. We suggest the same ideas used in the extraction of concepts to be applied in the method that we have proposed.

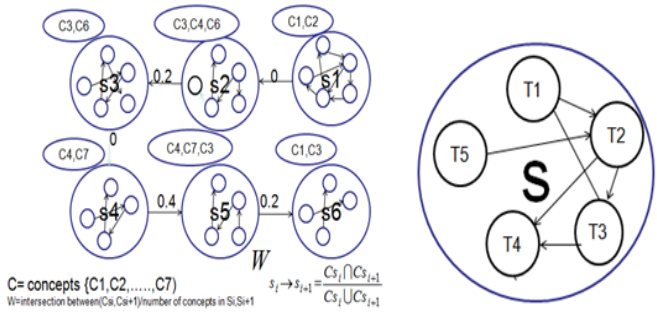

Fig. 1. Represent sentences in the nodes and extract the concepts from the sentences and the weight between the nodes refer to similarity between the sentences, T1, T2,... ,Tn represent the terms.

## 5 A NEW METHOD TO DETECT PLAGIARISM USING A GRAPH BASED DOCUMENT REPRESENTATION

In this section we discuss the new method for plagiarism detection based on the graph representation. The method relies on a number of steps, first we break down the document into its constituent sentences. Preprocessing for each document is required such as segmentation of each sentence into separated terms and stop word removal and. the stemming process is applied on the sentence, then we represent the sentences in the form of a nodes related to edge on the order of the sentence within the document. Where each node contains a one sentence of the document. To represent the terms of each sentence as graph we use the method mentioned in section 3. Where each node in the graph contains one term. The node are connected to each other according to order of term position in the sentence. The whole document consists of a number of nodes determined by the number of sentence in the document. Each node is a graph which represents a sentence. The concepts of node terms are extracted and used for calculating the similarity between each pair of nodes using e.q. (1).

$$W_{s_i \to s_{i+1}} = \frac{Cs_i \cap Cs_{i+1}}{Cs_i \cup Cs_{i+1}} ----\to (1)$$

Where $S_i$ is sentence 1 and $S_{i+1}$ is sentence following the sentence1 , $C_{si}$ is a number of concepts in sentence I and $C_{si+1}$ is a number of concepts in sentence i+1, W is a weight or similarity between $s_i$ and $s_{i+1}$.
After getting all the concepts of sentences those concepts grouped in one node called the topic signature, this node inked to each node in the graph and then the similarity between the topic signature and each node is computes separately, We calculate the similarity between the node of the topic signature and the other nodes based on shared concepts using the following equation:

$$W_{Topic\ Signature} = \frac{C_{s_i}}{C_s} ----\to (2)$$

Where $C_{si}$ is a number of concepts in the sentence I, C is the number of concepts in the topic signature node.
Topic signature node is formed by extracting the concepts of each sentence terms and grouping them in such node. The main advantage of the proposed method is the topic signature which is main entry for the graph is used as quick guide to the relevant nodes. which should be considered for the comparison between source documents and suspected one. For example, if there is a matching between concept1 on the suspected document and concept 1 in the original document, we go directly to the nodes that containing of concept1 in both original document and the suspected document and we ignore all the remaining sentences. In the case of full matching of topic signature of original document with topic signature of suspected document, we will face the problem of the huge number of comparisons which is taken as disadvantage in current methods. To avoid such problem, only we compare the most important nodes. To determine those important nodes we link each node with all the nodes, However, each node has a number of in-links and out-links then compute the similarity of each sentence with the rest of the sentences in the document. Based on the similarities, we extract the highest degree of similarity between the nodes which define the most important nodes and ignore the rest. To calculate the degree of similarity of the node with other nodes using the following equations

$$W_{out-link(si \to sk)} = \frac{c_{si} \cap c_{sk}}{c_{sk}} ----\to (3)$$

$$W_{in-link(si \to sk)} = \frac{c_{si} \cap c_{sk}}{c_{si}} ----\to (4)$$

Where is a sentence I, $C_{si}$ is a number of concepts in the sentence I, $C_{sk}$ is a number of concepts in each sentence in the graph.

## 6 CONCLUSION

In this paper we have considered the problem of plagiarism detection, one of the most publicized forms of text reuse around us today. In particular, we have focused on plagiarism detection using graph based document representation. We have discussed various approaches of plagiarism detection. To date there are few resources which specifically address the task of plagiarism detection.

The proposed method to detect the plagiarism based on graph representation required break down the document into its constituent sentences and the graph building by grouping each sentence terms in one node, the resulted nodes are connected to each other based on order of sentence within the document, all nodes in graph are also connected to top level node" Topic Signature ".

Topic signature is a main entry for the graph is used as quick guide to the relevant nodes. Which should be considered for the comparison between source documents and suspected one.

The proposed method based on graph representations contributed by increase the efficiency and reduce the huge a number of matching process.

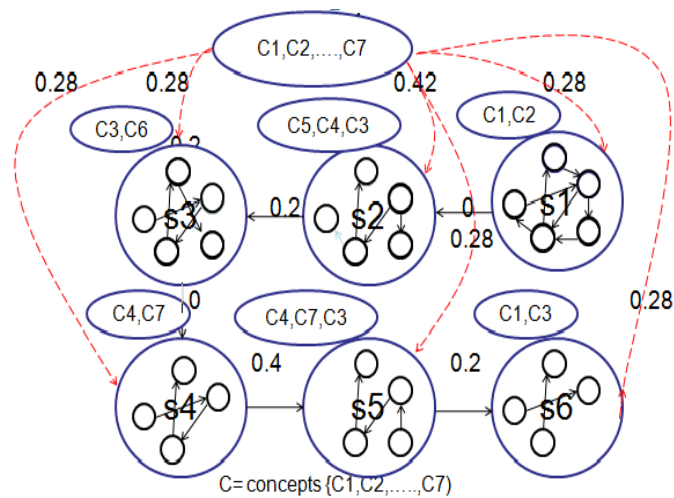

*Fig. 2. Topic Signature formulation for concepts Sentence as Graph*

**Mr. Ahmed Hamza** is a PHD student in Universiti Teknologi Malaysia. Presently working as a lecturer in the Faculty of Computer Science in International University of Africa(IUA) in Sudan .I have bachelor degree in Computer Science from IUA in 2004. I received master degree in Computer Science from Sudan University of science and technology in 2008. I current research interest includes Information Retrieval, Database, and Data mining.

**Dr. Naomie Salim** is an Associate Professor presently working as a Deputy Dean of Research & Postgraduate Studies in the Faculty of Computer Science and Information System in Universiti Teknologi Malaysia. She received her bachelor degree in Computer Science from Universiti Teknologi Malaysia in 1989. She received her master degree in Computer Science from University of West Michigan in 1992. In 2002, she received her Ph.D (Computational Informatics) from University of Sheffield, United Kingdom. Her current research interest includes Information Retrieval, Distributed Database and Chemoinformatic.

**Mr. Mohammed Salem Binwahlan** received his B.Sc. dgree in Computer Science from Hadhramout University of Science and Technology, Yemen in 2000. He received his Master degree from Universiti Teknologi Malaysia in 2006. He is currently with Hadhramout University of Science and Technology as lecturer and pursuing Ph.D degree in the Faculty of Computer Science and Information System, Universiti Teknologi Malaysia. His current research interest includes Information Retrieval, Text Summarization and Soft Co puting.